\documentclass{PHYEAUTH}
\usepackage{graphicx}
\usepackage{amsmath}
\usepackage{amssymb}

\begin{document}

\begin{frontmatter}



\title{Exchange and correlation effects on drag in low density electron bilayers:
Coulomb and virtual-optical-phonon-mediated electron-electron interaction}


\author[chonnam]{S. M. Badalyan\corauthref{cor1}},
\author[chonnam]{C. S. Kim}, and \author[columbia]{G. Vignale}

\corauth[cor1]{Corresponding author. Also at: Department of
Radiophysics, Yerevan State University, 1 A. Manoogian st., 375025
Yerevan, Armenia. Phone: 8262-530-0063, Fax: 8262-530-3369, Email
address: badalyan@boltzmann.chonam.ac.kr}

\address[chonnam]{Department of Physics and Institute for Condensed
Matter Theory, Chonnam National University, Gwangju 500-757,
Korea}
\address[columbia]{Department of Physics and Astronomy, University of
Missouri - Columbia, Missouri 65211, USA}

\begin{abstract}
We  investigate the effect of exchange and correlation (xc) in
low-density electron bilayers. Along with the direct Coulomb
interaction, the effective electron-electron interaction mediated
by the exchange of virtual polar optical (PO) phonons is
considered. We find that the introduction of xc corrections
results in a significant enhancement of the transresistivity and
qualitative changes in its temperature dependence. The virtual
PO-phonon contribution behaves similarly to the Coulomb drag and
reduces noticeably the total drag thereby resulting in a better
agreement with the recent experimental findings.
\end{abstract}

\end{frontmatter}

\section{Introduction}

Recently the drag measurements have been extended to the limit of
very low carrier density. The dimensionless parameter
$r_{s}=\sqrt{2}/(k_{F}a_{B}^{\ast })$, which is used to describe
the carrier density, $n$, and measures the strength of the
electron-electron (e-e) interaction \cite{GV05}, varies
approximately from $10$ to $20$ in the experiment on hole samples
\cite{pillar}. Here $k_{F}=\sqrt{2\pi n}$ is the Fermi wave
vector, $a_{B}^{\ast}$ the effective Bohr radius. In electron
samples $r_{s}$ is appreciably smaller \cite{kellogg}, owing to
the small electron mass. However, it is clear from the dimensional
analysis that in both types of low-density samples, the Coulomb
potential energy dominates the kinetic energy, and an adequate
description of the drag cannot be provided by simple
Boltzmann-equation theories \cite{jauho}, which do not include the
strong exchange-correlation (xc) effects.

The interaction effects on drag have been addressed previously in
several experimental \cite{hill,noh1} and theoretical
\cite{flensberg,swierkowski} papers. While the theoretical
prediction \cite{flensberg} of an enhancement of drag by plasmons
has been experimentally verified \cite{hill} in high density
electron samples, important differences have been reported
\cite{noh1} from the results obtained within the random phase
approximation (RPA). Lately, motivated by the recent experiments
\cite{pillar,kellogg}, the transresistivity has been calculated in
Refs.~\cite{hwang} and \cite{yurt}. Both works have included only
the exchange effects in the static limit via $q$-dependent but
$\omega$-independent local field factors (LFF). Besides, in the
adopted approximation the nondiagonal inter-layer LFF have been
taken to be zero while for the intra-layer LFF the simple Hubbard
approximation has been used, which significantly underestimates
the LFF.

We propose a new approach making use of the dynamic xc kernels to
study the xc effects on the drag in electron bilayers. The
introduction of xc results in a significant enhancement of the
drag rate and qualitative changes in its temperature dependence.
In particular, a large high-temperature plasmon peak that is
present in the RPA disappears when the xc corrections are
included. We combine the direct Coulomb interaction with an
effective e-e interaction, mediated by the exchange of virtual
polar optical (PO) phonons. This new contribution to the drag
improves noticeably a reasonably good agreement of our numerical
results with the experimental findings by Kellogg \textit{et
al.}~\cite{kellogg}.

\section{Theoretical formulation}
\begin{figure}[t]
\centering
\includegraphics[width=5.5cm,angle=-90]{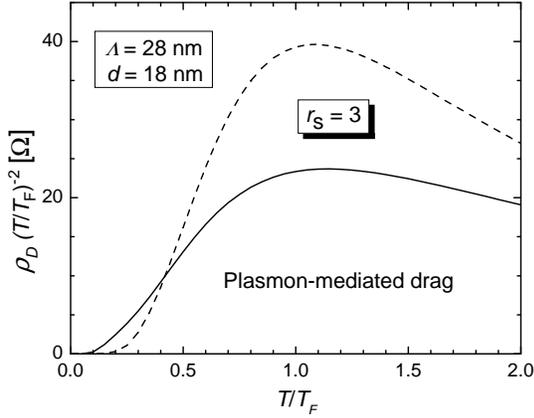}
\caption{Dynamic xc corrections to the plasmon-mediated drag for
$r_{s}=3 $. The scaled transresistivity vs temperature is shown
within the RPA (dashed curve) and beyond it including the intra-
and inter-layer xc effects (solid curve).} \label{fig0}
\end{figure}

\begin{figure}[t]
\centering
\includegraphics[width=5.5cm,angle=-90]{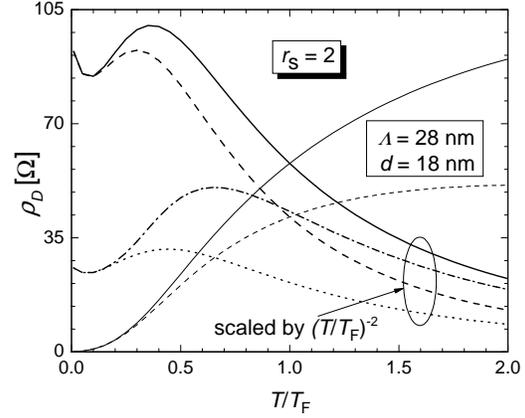}
\caption{The drag transresistivity vs temperature for $r_{s}=2$.
The solid and dashed (dash-dotted and dotted) curves represent the
total $\rho _{_{D}} $ and the particle-hole contribution to $\rho
_{_{D}}$ with the xc corrections included (within the RPA).}
\label{fig1}
\end{figure}

The drag transresistivity is given by \cite{rojo}
\begin{eqnarray}
\rho_{_{D}}&=&{\frac{\hbar^{2}}{2e^{2}n_{1}n_{2}TA}}
\sum_{\overrightarrow{q}}q^{2}\int_{0}^{\infty}\frac{d\omega}{2\pi}
\left\vert W_{12}(q,\omega )\right\vert^{2} \nonumber \\
&\times& \frac{{\text{Im}}\Pi _{1}^{0}(q,\omega ){\text{Im}}\Pi
_{2}^{0}(q,\omega )}{\sinh ^{2}(\hbar \omega /2T)}\label{eq1}
\end{eqnarray}
where $A$ is the normalization area, $\Pi _{1,2}^{0}(q,\omega )$
the finite temperature electron polarization function of
individual layers in the absence of inter-particle interaction.
The xc effects on the dynamically screened inter-layer e-e
interaction is approximated by $W_{12}(q,\omega
)=V_{xc,12}(q,\omega )/\varepsilon _{{bi}}(q,\omega )$ with
$\varepsilon _{{bi}}(q,\omega )=\varepsilon _{1}(q,\omega
)\varepsilon _{2}(q,\omega )-V_{xc,12}(q,\omega )^{2}\Pi
_{1}^{0}(q,\omega )\Pi _{2}^{0}(q,\omega )$ being the bilayer
screening function \cite{swierkowski,zheng1}. Here we introduce
the screening functions $\varepsilon _{1,2}(q,\omega )$ of
individual layers. The intra- and inter-layer unscreened effective
Coulomb interactions are given by
\begin{equation}
V_{{xc},ij}^{C}(q,\omega )=v(q)\left( 1-G_{{xc,}ij}(q,\omega )\right)
F_{ij}(qd,q\Lambda )  \label{veff}
\end{equation}
where the intra- and inter-layer LFF, $G_{{xc},ij}(q,\omega )$,
decrease effectively the bare Coulomb interaction, $v(q)$, by a
factor of $1-G_{{xc,}ij}(q,\omega )$. The form factors
$F_{ij}(qd,q\Lambda )=\int dz dz^{\prime }\rho _{i}(z)\rho
_{j}(z^{\prime })\exp (-q\left\vert z-z^{\prime }\right\vert )$
are obtained making use $\rho (z)=(2/d)\,\sin (\pi z/d)^{2}$ for
the electron density profiles. Here $\Lambda $ is the inter-layer
spacing and $d$ the width of quantum wells.
%The RPA is recovered if $G_{{xc},ij}(q,\omega )$ are set to zero.

\begin{figure}[t]
\centering
\includegraphics[width=5cm,angle=-90]{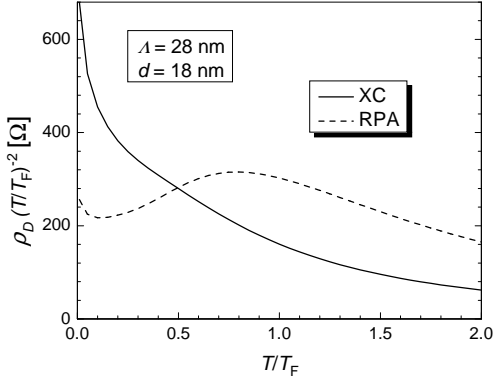}
\caption{The full xc corrections to the drag for $r_{s}=3 $. The
dashed and solid curves represent the scaled transresistivity vs
temperature, calculated, respectively, within and beyond the RPA.
The RPA data are multiplied by a factor of $3$.} \label{fig2}
\end{figure}

The phonon-mediated effective unscreened e-e interaction,
$V_{{xc},ij}^{ph}(q,\omega )$, appears in second order
perturbation theory with respect to the bare electron-phonon (e-p)
coupling. Neglecting the weak energy dispersion of PO-phonons, we
find that $V_{{xc},ij}^{ph}(q,\omega )$ is proportional to the
PO-phonon propagator $D_{{PO}}(\omega )={2\hbar ^{-1}\omega
}_{{PO}}/{\left( \omega ^{2}-\omega _{{PO}}^{2}\right) }$ and is
represented as
\begin{equation}
V_{{xc},ij}^{ph}(q,\omega )=\left(\frac{\kappa
_{0}}{\kappa_{\infty }}-1\right) \frac{{\hbar }\omega
_{PO}}{2}D_{{PO}}(\omega )V_{{xc},ij}^{C}(q,\omega )
\end{equation}
where $\kappa _{0}$ and $\kappa _{\infty }$ are the static and
high frequency dielectric constants. We are interested in
experimental situations where $T$ is much lower than the PO-phonon
energy, $\hbar \omega _{_{PO}}\approx 421$ K. Due to the energy
and momentum conservation, the real PO-phonon emission and
absorption processes are accompanied with a large energy and
momentum transfer. Therefore, the resulting drag is negligibly
small. This is clearly seen in Ref.~\cite{opt} where the
real-PO-phonon contribution to drag is obtained by a factor of
$e^{-28}$ smaller than the direct Coulomb drag. In stark contrast,
scattering by virtual PO-phonons does not imply the energy and
momentum conservation and provides an appreciable contribution to
drag. Moreover, the main contribution to drag is made by
scattering processes with the energy scale of $\omega \lesssim
T\ll \omega _{PO}$ for which the PO-phonon propagator
$D_{PO}(q,\omega )$ is far from the mass surface. Thus, the total
unscreened e-e interaction, $V_{{xc},ij}(q,\omega
)=V_{{xc},ij}^{C}(q,\omega )+V_{{xc},ij}^{ph}(q,\omega )$, is well
approximated by
\begin{equation}
V_{{xc},ij}(q,\omega )=\left( 2-\kappa _{0}/\kappa _{\infty
}\right) V_{{xc},ij}^{C}(q,\omega ).
\end{equation}
It is seen that the attractive e-e interaction, mediated by
virtual PO-phonons, reduces the Coulomb interaction strength by a
factor of $\left( 2-\kappa _{0}/\kappa _{\infty }\right) $. In
GaAs taking $\kappa _{0}=13.18$ and $\kappa _{\infty }=10.89$
\cite{adachi}, we have an about $21$ percent reduction of the
interaction strength. Notice that the reduction effect manifests
itself also via the screening of interaction. The acoustical
phonons, because of their weak e-p coupling and small energy with
the strong linear dispersion, have no such universal effect on
drag and are not considered here.

To obtain the bilayer LFF $G_{{xc},ij}(q,\omega )$ we interpret
the layer index as an isospin\ and represent the intra- and
inter-layer LFF in terms of the "spin-channel" and
"charge-channel" LFF. When treating the plasmon contribution to
the drag, we employ the dynamic spin-spin and charge-charge xc
kernels, evaluated by Qian and Vignale \cite{qv}. The
frequency-dependence of the LFF in the particle-hole continuum
region (finite wave vector, low frequency) is still largely
unknown. For this reason, in evaluating the particle-hole
contribution to the drag, we use the static limit of the LFF
taking advantage of the analytical expressions recently obtained
for the spin-spin and charge-charge LFF \cite{davoudi}.

\section{Many-body corrections to the total drag transresistivity}
\begin{figure}[t]
\centering
\includegraphics[width=5cm,angle=-90]{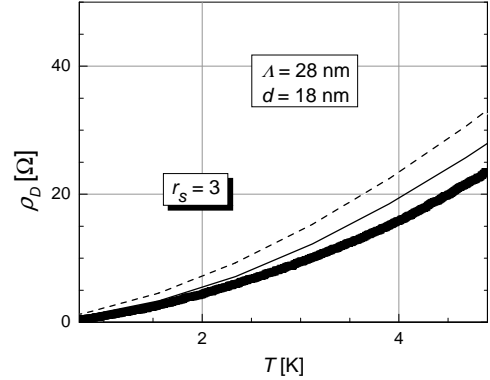} \caption{The
xc effect on the total drag (solid curve) for $r_{s}=3$ due to the
combined effect of direct Coulomb (dashed curve) and effective
virtual-PO-phonon-mediated e-e interaction. The symbols are the
experimental findings from Ref.~\cite{kellogg} for $n=3.8\cdot
10^{10}$ cm$^{-2}$. } \label{fig3}
\end{figure}

In Fig.~\ref{fig0} we show the effect of the intra- and
inter-layer dynamic xc on the plasmon-mediated drag by comparing
the obtained results with those of RPA-based calculations for the
symmetric bilayer systems with $r_{s}=3$. It is seen that the
upturn temperature in the scaled transresistivity vs temperature
essentially decreases when the xc effects are included. The
plasmons begin to contribute heavily at temperatures about
$0.1T_{F}$, which is approximately a factor of $2$ smaller than
the upturn temperature calculated within the RPA. The effect is
slightly weak for $r_{s}=2$. At the relatively high temperatures,
the average energy of plasmons that mediate the drag increases.
Hence, the plasmon damping, caused by dynamic xc effects, becomes
stronger with a consequent reduction in the drag transresistivity.
Thus, around $0.4T_{F}$ the plasmon-mediated drag obtained within
the RPA exceeds the drag which takes into account the full xc
effects. Notice also that at even higher temperatures the
differences in the magnitude of drag in the different
approximations diminishes with increasing $T$.

In Fig.~\ref{fig1} we plot the drag transresistivity vs
temperature within and beyond the RPA for $r_{s}=2$. It is seen
that in both approximations the particle-hole contribution to the
scaled drag transresistivity first shows a slight dip followed by
a peak at higher temperatures. In the RPA the plasmon contribution
to drag enhances this peak and shifts it to even higher
temperatures so that the total transresistivity shows a pronounced
peak approximately at the position where the plasmon contribution
has a peak. The introduction of the static exchange-controlled LFF
increases the peak height of the particle-hole contribution to the
drag and shifts the peak towards lower temperatures. On the other
hand, the plasmon-mediated drag is moderately suppressed by the
dynamic xc corrections (cf. Fig.~\ref{fig0}). Thus, the resulting
peak in the graph of the total transresistivity vs temperature
remains a small feature at relatively low temperatures, while at
high temperatures the drag rate shows a monotonic decrease in $T$.

The described quantitative and qualitative differences in the
behavior of the scaled total transresistivity within and beyond
the RPA becomes more pronounced at lower densities. As seen from
Fig.~\ref{fig2} for $r_{s}=3$ the total transresistivity beyond
the RPA as a function of temperature shows no peak at all and this
is in stark contrast to the peaked behavior of the
transresistivity within the RPA. The disappearance of the large
high-temperature plasmon peak results from the strong increase of
the drag transresistivity at low temperatures: we ascribe this to
the fact that the contribution to drag, made by large-angle
inter-layer scattering processes, becomes dominant when the
many-body xc corrections are included. The large-angle scattering
component strongly enhances the drag for two reasons. First, at
low $T$ the e-e scattering phase-space diverges near $x\equiv
q/2k_{F}=1$. On the other hand, when $x\simeq 1$, the static
intra-layer LFF $G_{11}(x)$ becomes close to unity, leading to a
reduction of the effective intra-layer interaction
$V_{{xc},11}(x)$. This by itself weakens the dynamic screening and
enhances the drag. Thus, on the background of the large
transresistivity at small $T$, the plasmon-mediated contribution
to drag does not result in a peaked behavior of the scaled
transresistivity and the total drag rate decreases monotonically
in $T$.

Lack of experimental measurements of the drag at high temperatures
does not allow for the time being an experimental verification of
our predictions on the position and strength of the plasmon peak
in low density bilayers. Notice however that the low temperature
peak of the transresistivity and its monotonous behavior at high
temperatures, observed in the experiment on hole samples in
Ref.~\cite{pillar}, by implication support our prediction. At low
temperatures, as shown in Fig.~\ref{fig3}, our numerical results
are in reasonably good agreement with the experimental findings by
Kellogg {\it et al.}~\cite{kellogg}. It is clear that the
introduction of the effective e-e interaction, mediated by the
exchange of virtual PO-phonons, reduces noticeably the total drag
and results in a better agreement with experiment.

We thank Zhixin Qian for discussions and kindly making the results
of his calculations accessible prior to publication. We
acknowledge support from the Korea Science and Engineering
Foundation Grant No.~R05-2003-000-11432-0 and NSF Grant No.
DMR-0313681.

\end{document}